\documentstyle[aps,twocolumn,epsfig]{revtex}
\begin{document}
%\preprint{SUNY-NTG-99-xx}
\newcommand{\beq}{\begin{equation}}
\newcommand{\eeq}{\end{equation}}
\newcommand{\bea}{\begin{eqnarray}}
\newcommand{\eea}{\end{eqnarray}}
\newcommand{\bfig}{\begin{figure}}
\newcommand{\efig}{\end{figure}}
\newcommand{\ie}{{\it i.e.}}
\newcommand{\bce}{\begin{center}}
\newcommand{\ece}{\end{center}}
\newcommand{\eg}{{\it e.g.}}
\newcommand{\etal}{{\it et al.}}
\def\lsim{\mathrel{\rlap{\lower4pt\hbox{\hskip1pt$\sim$}}
    \raise1pt\hbox{$<$}}}         %less than or approx. symbol
\def\gsim{\mathrel{\rlap{\lower4pt\hbox{\hskip1pt$\sim$}}
    \raise1pt\hbox{$>$}}}         %greater than or approx. symbol
\twocolumn[\hsize\textwidth\columnwidth\hsize\csname @twocolumnfalse\endcsname
\title
%{$\rho$-Mesons from $\pi^+\pi^-$ Decays in Dilute Hadronic Matter}
{$\pi^+\pi^-$ Emission in High-Energy Nuclear Collisions}
 
\author
{Ralf Rapp}

\address{
 NORDITA, Blegdamsvej 17, DK-2100 Copenhagen, Denmark 
% \\  2) 
   }
 
\maketitle
 
\begin{abstract}
Realistic vacuum $\pi\pi$ interactions are employed to investigate
thermal $\pi^+\pi^-$ emission spectra from the late stages of heavy-ion
reactions at ultrarelativistic energies. Hadronic in-medium effects,
including many-body $\rho$-meson spectral functions used earlier to
describe the dilepton excess observed at CERN-SPS energies, are 
implemented to assess resulting modifications in relation to 
recent measurements of $\pi^+\pi^-$ invariant-mass spectra by the STAR 
collaboration in $p$-$p$ and $Au$-$Au$ collisions at 
$\sqrt{s_{NN}}$=200~GeV. Statistical model estimates for the 
$\rho^0/\pi^-$ and $K^*/K$ ratios close to the expected thermal
freezeout are also given. 
\end{abstract}

]
 
%%%%%%%%%%%%%%%%%%%%%%%%%%%%%%%%%%%%%%%%%%%%%%%%%%%%%%%%%%%%%%%%%%%%%%%%
\section{Introduction}
%%%%%%%%%%%%%%%%%%%%%%%%%%%%%%%%%%%%%%%%%%%%%%%%%%%%%%%%%%%%%%%%%%%%%%%%

The unambiguous identification of hadron modifications in hot and/or 
dense matter constitutes one of the major objectives in modern nuclear 
physics~\cite{BR02a,RW00}. Hadronic states whose vacuum properties are 
related to the spontaneous breaking of chiral symmetry, are
of particular interest as changes in their spectral distribution 
reflect precursor phenomena towards the chiral phase transition. 
A versatile environment for these investigations is provided by 
high-energy heavy-ion collisions, as different center-of-mass 
({\em CM}) energies have been shown~\cite{BRS03} to probe large regions
in temperature and density of the QCD phase diagram.  

On the one hand, dilepton observables~\cite{ceres95,helios3,ceres02}
have been proved to be powerful tools to extract medium modifications of 
vector-mesons~\cite{RW00} during the evolution
history of the hot and dense fireballs formed in heavy-ion reactions. 
However, 
%large nonresonant background sources together with 
the relative signal weakness (parametrically suppressed by $\alpha^2$ 
compared to hadronic sources) entails appreciable experimental 
uncertainties, which, as of now, do not allow for a unique theoretical 
interpretation of the results. 
On the other hand, hadronic final state measurements of strongly 
decaying resonances have thus far not revealed  significant medium 
effects~\cite{e917,na49}. This is believed to be mostly due to the 
diluteness of the matter from which the decay products are able to 
reach the detectors undistorted (unlike in the dilepton case). In this 
context, recent (preliminary) data~\cite{star03} on $\pi^+\pi^-$ 
invariant-mass distributions from $\sqrt{{s}_{NN}}$=200~GeV nuclear 
collisions by the STAR collaboration are of particular 
interest. Assuming Breit-Wigner parametrizations of the underlying 
resonance structures ($\rho$, $\omega$, $f_0$, etc.),  a downward
shift of $\sim$~30~MeV for the peak position in the $\pi^+\pi^-$ 
distribution from $\rho$-meson decays has been 
reported~\cite{star03} when going from $p$-$p$ to 40-80\% central 
$Au$-$Au$ collisions.    
In addition, a shift of -40~MeV relative to
a nominal $\rho$-meson mass of 770~MeV has been extracted from the 
$p$-$p$ data.   

Two recent articles have specifically addressed these observations (see
also Ref.~\cite{BBFSB91} for an earlier study).  
In Ref.~\cite{SB03}, a number of hadronic $s$- and $t$-channel exchanges 
between $\rho$-mesons and the surrounding matter close to freezeout have 
been estimated to produce a $\sim$50~MeV downward mass shift, mostly 
originating from scalar $t$-channel exchanges with anti-/baryons (in 
line with "Brown-Rho" scaling~\cite{BR02}).  
In Ref.~\cite{KP03} a more generic analysis of the interplay between 
phase-space, mass-shift and width effects has been carried out, with 
the latter two implemented via schematic selfenergies in the 
underlying $\rho$-meson spectral function. 

In the present article we go beyond these analyses in several respects.
We will assess $\pi^+\pi^-$ spectra based on microscopic $\pi\pi$ 
interactions that accurately describe pertinent free scattering phase 
shifts in $S$-, $P$- and $D$-waves, including hadronic medium 
modifications through finite-temperature effects on intermediate 
two-pion states.  For the $\rho$ meson we will also use more advanced
in-medium many-body spectral functions which have been constructed and 
applied in recent years mostly in the context of dilepton production 
(with fair success, \eg, in reproducing the low-mass excess observed by 
CERES/NA45), cf.~Ref.~\cite{RW00} for a recent review. 
This is critical for illuminating the important issue to what extent 
the relative apparent 
mass shift of about 30~MeV between $p$-$p$ and $Au$-$Au$ experiments 
is consistent with such an approach, and what responsible 
mechanisms are. Another question that will be addressed concerns
total $\rho$-meson yields, \ie, the $\rho^0/\pi^-$ ratio and   
its evolution from chemical to thermal freezeout.  
%We will furthermore elaborate on differences between hadronic 
%($\pi^+\pi^-$) and electromagnetic ($e^+e^-$) probes of the $\rho$ 
%spectral function in heavy-ion collisions.     
 
The article is organized as follows. 
In Sec.~\ref{sec_rho} we apply microscopic in-medium $\rho$ 
spectral functions to compute thermal $\pi^+\pi^-$ emission rates
at RHIC, including a simple evolution model to assess  
uncertainties due to freezeout conditions. In Sec.~\ref{sec_SDwave} 
we establish a relation between thermal emission rates and 
$\pi\pi$ scattering amplitudes, and employ a realistic $\pi\pi$
interaction with (thermal) in-medium effects to evaluate 
S-wave ("$\sigma$") and $D$-wave ($f_2(1270)$) 
contributions to the emission spectra.  
In Sec.~\ref{sec_feed} we investigate both spectral shape
and magnitude of resonance feeddown to the $\rho$-meson yield, 
as well as its evolution from chemical to thermal freezeout in
central heavy-ion collisions. We also comment on the related 
behavior of the recently measured $K^*/K$ ratio in the hadronic 
evolution. 
In Sec.~\ref{sec_concl} we summarize and conclude.

%%%%%%%%%%%%%%%%%%%%%%%%%%%%%%%%%%%%%%%%%%%%%%%%%%%%%%%%%%
\section{In-Medium $\rho$-Mesons and $\pi^+\pi^-$ Emission}
\label{sec_rho}
%%%%%%%%%%%%%%%%%%%%%%%%%%%%%%%%%%%%%%%%%%%%%%%%%%%%%%%%%%

%%%%%%%%%%%%%%%%%%%%%%%%%%%%%%%%%%%%%%%%%%%%%%%%%%%%%%%%%%%%%%%%%%%%%%%%
\subsection{The $\rho$-Meson in Hot Hadronic Matter} 
\label{ssec_in-med-rho}
%%%%%%%%%%%%%%%%%%%%%%%%%%%%%%%%%%%%%%%%%%%%%%%%%%%%%%%%%%%%%%%%%%%%%%%%
Let us briefly recall the main elements in the calculation
of the medium-modified $\rho$-meson spectral function
along the lines of our previous work~\cite{RCW97,UBRW98,RW99}. 
At finite temperatures and baryon densities, the  
$\rho$-propagator can be cast into the form 
\beq
D_\rho=\frac{1}{M^2-(m_\rho^{(0)})^2-\Sigma_{\rho\pi\pi}
-\Sigma_{\rho M} -\Sigma_{\rho B} } \ .
\label{Drho}
\eeq
The in-medium selfenergy insertions consist of three parts: 
(i) $\Sigma_{\rho\pi\pi}$ encodes the free decay width into 2-pion 
states which in matter is modified by standard "pisobar" excitations, 
$\Delta N^{-1}$ and $NN^{-1}$~\cite{AKLQ92,CS93,HFN93,RCW97}, 
extended to finite total 3-momentum~\cite{UBRW98} as well as finite 
temperatures (most importantly pion Bosefactors), see also
Refs.~\cite{Hees00,UBW02}.
Effects of pion excitations into higher resonances, as well as
interactions with hyperons, $N^*$- and $\Delta^*$-states, are 
approximated by using an "effective" nucleon density 
$\varrho_N^{eff}= (\varrho_N+\frac{1}{2}\varrho_{B^*})$~\cite{RW99}; 
(ii) $\Sigma_{\rho M}$ describes resonant $\rho$-interactions with 
surrounding $\pi$, $K$ and $\rho$ mesons (for resonances with 
established decays into $\rho$'s up to masses 
$\sim$1.6~GeV)~\cite{RG99,EK99};  
(iii) $\Sigma_{\rho B}$ accounts for the resonant $\rho$-interactions
with surrounding nucleons, hyperons and baryon 
resonances~\cite{FP97,RCW97,PPLLM,RW99}.  
All underlying hadronic vertices (characterized by a coupling constant
and hadronic formfactor cutoff) have been constrained by both hadronic
and radiative decay branchings. In addition, baryonic
contributions have been subjected to a comprehensive fit to 
photoabsorption data on nucleons and nuclei~\cite{RUBW}. This, in 
particular, leaves no room for additional $t$-channel exchanges.
%(which in Ref.~\cite{SB03} are the main source of the $\rho$-mass 
%shift close to freezeout under RHIC conditions).  

At collider energies (RHIC and LHC) the {\em net} baryon densities
at midrapidity are rather small. However, as pointed out in 
Ref.~\cite{Ra01}, $CP$-invariance of strong interactions implies 
that the $\rho$ meson equally interacts with baryons and anti-baryons.
Thus, the relevant quantity for medium effects is the {\em sum} of 
baryon and antibaryon densities, \ie, 
$\varrho_{B+\bar B}=\varrho_B+\varrho_{\bar B}\simeq 
\varrho_B * (1+\bar p/p)$, 
where $\bar p/p$ denotes the experimentally measured ratio
of antiprotons to protons.  

The main medium modification of the resulting $\rho$ spectral function,
$A_\rho\equiv-2{\rm Im}D_\rho$, consists of a substantial broadening 
of its resonance shape~\cite{RW99,Ra01}, even at moderate temperatures 
and, more importantly, baryon densities. The pole position, on the 
other hand, is affected rather little. This is due to the quite generic 
feature that many-body contributions to the real part of the selfenergy 
are of varying sign and thus tend to cancel, whereas the imaginary parts 
are of definite sign and thus strictly add up.

%%%%%%%%%%%%%%%%%%%%%%%%%%%%%%%%%%%%%%%%%%%%%%%%%%%%%%%%%%%%%%%%%%%%%%%
\subsection{$\rho\to \pi\pi$ Decays at Freezeout}
\label{ssec_rhopipi}
%%%%%%%%%%%%%%%%%%%%%%%%%%%%%%%%%%%%%%%%%%%%%%%%%%%%%%%%%%%%%%%%%%%%%%%
Following Ref.~\cite{Weld93}, the thermal rate of two-pion emission
from neutral $\rho$ decays per unit four-volume and four-momentum can
be written as
\bea
\frac{dN_{\rho^0\to\pi^+\pi^-}}{d^4qd^4x} &=& \frac{3}{4\pi^4} \ 
f^\rho(q_0;\mu_\rho,T) \ {\rm Im} D_\rho(M,q;\mu_B,T)  
\nonumber\\  
 & & \times \ {\rm Im} \Sigma_{\rho\pi\pi}(M) \ , 
\label{dNdxdq}
\eea
where $f^\rho$ denotes the usual Bosefactor (with 
chemical potential $\mu_\rho$), and Im$D_\rho$  
depends on temperature, baryon-chemical 
potential $\mu_B$ and possibly other (meson-) chemical potentials.   
If the emitted pions reach the detector 
undistorted, the imaginary part of the two-pion selfenergy
is given by the free decay width, 
$-{\rm Im} \Sigma^{vac}_{\rho\pi\pi}(M) 
= m_\rho \Gamma^{vac}_\rho(M)$, 
which depends on the invariant $\rho$-mass $M$ only.  
Using the standard $\rho\pi\pi$ interaction vertex, 
\beq
{\cal L}_{\rho\pi\pi}= g_{\rho} \ {\vec \rho}^\mu \cdot
         (\vec \pi \times \partial_\mu \vec\pi) \ ,                     
\eeq
its explicit form is (cf., \eg, Refs.~\cite{CS93,RCW97})
\beq
{\rm Im} \Sigma_{\rho\pi\pi} = -\frac{g_\rho^2}{6\pi} \ \frac{k^3}{M} 
 \ F_{\rho\pi\pi}(k^2)^2 \ , 
\label{ImSig}
\eeq
with $F_{\rho\pi\pi}$ a (weakly momentum-dependent) hadronic formfactor
(which suppresses  the selfenergy at large masses $M$),
and $k=(M^2/4-m_\pi^2)^{1/2}$ the pion decay 3-momentum. 
%Even without rescattering, the emitted two pions can still sense
%their environment through Bose-Einstein correlations cf., \eg, 
%Ref.~\cite{Laf93}, which amonts to introducing a factor 
%$[1+2f^\pi(\omega_k)]$ into the selfenergy, Eq.~(\ref{ImSig}).  
%We will inclide this correction below.  
An invariant-mass distribution is obtained upon integrating 
Eq.~(\ref{dNdxdq}) over the thermal $\rho$ 3-momentum 
distribution\footnote{if momentum acceptance cuts on the detected
pions are imposed, the 3-momentum integral in Eq.~(\ref{dNdxdM}) 
is subject to nontrivial corrections which will then also depend, 
\eg, on the collective flow velocity $\rho$-meson at the moment of
decay.}, 
\bea
\frac{dN_{\rho\to\pi\pi}}{d^4x dM} &=& \frac{6}{\pi} \ 
 {\rm Im}\Sigma_{\rho\pi\pi}(M) \ M \int \frac{d^3q}{q_0 (2\pi)^3} 
 \ f^\rho(q_0;\mu_\rho,T) 
\nonumber\\
& & \times \ {\rm Im} D_\rho(M,q;\mu_B,T) \ .    
\label{dNdxdM}
\eea
If one neglects the 
(typically weak~\cite{RW99,UBRW98}) 3-momentum dependence of the 
in-medium $\rho$ spectral function, and invokes the nonrelativistic
limit for the $\rho$-meson (scalar) density~\cite{KP03}, one arrives 
at ($R\equiv dN/d^4x$)
\bea
\frac{dR_{\rho\to\pi\pi}}{dM} &=& -\frac{6}{\pi} \ 
\frac{g_\rho^2}{6\pi} \ \frac{k^3}{M} \ F_{\rho\pi\pi}(k)^2 \ 
%[1+2f^\pi(\omega_k)]  
\nonumber\\
 & & \times  {\rm Im} D_\rho(M;\mu_B,T) \ 
\left(\frac{MT}{2\pi}\right)^{\frac{3}{2}} \ 
{\rm e}^{-(M-\mu_\rho)/T} \ . 
\nonumber\\
\label{dRdM}
\eea  
The above expression differs from the corresponding one in 
Ref.~\cite{KP03} by (i) an additional $1/M$-factor (as well as
formfactor) in the vacuum width, 
%(ii) the final state Bosefactor,
and (ii) the microscopic spectral function, which, most notably, 
leads to an additional $M$-dependence in Im$\Sigma_\rho^{tot}$ 
(appearing in the numerator of the spectral function), which is 
reminiscent (but not equal) to the free selfenergy.  
 
\bfig[htb]
\bce
\epsfig{figure=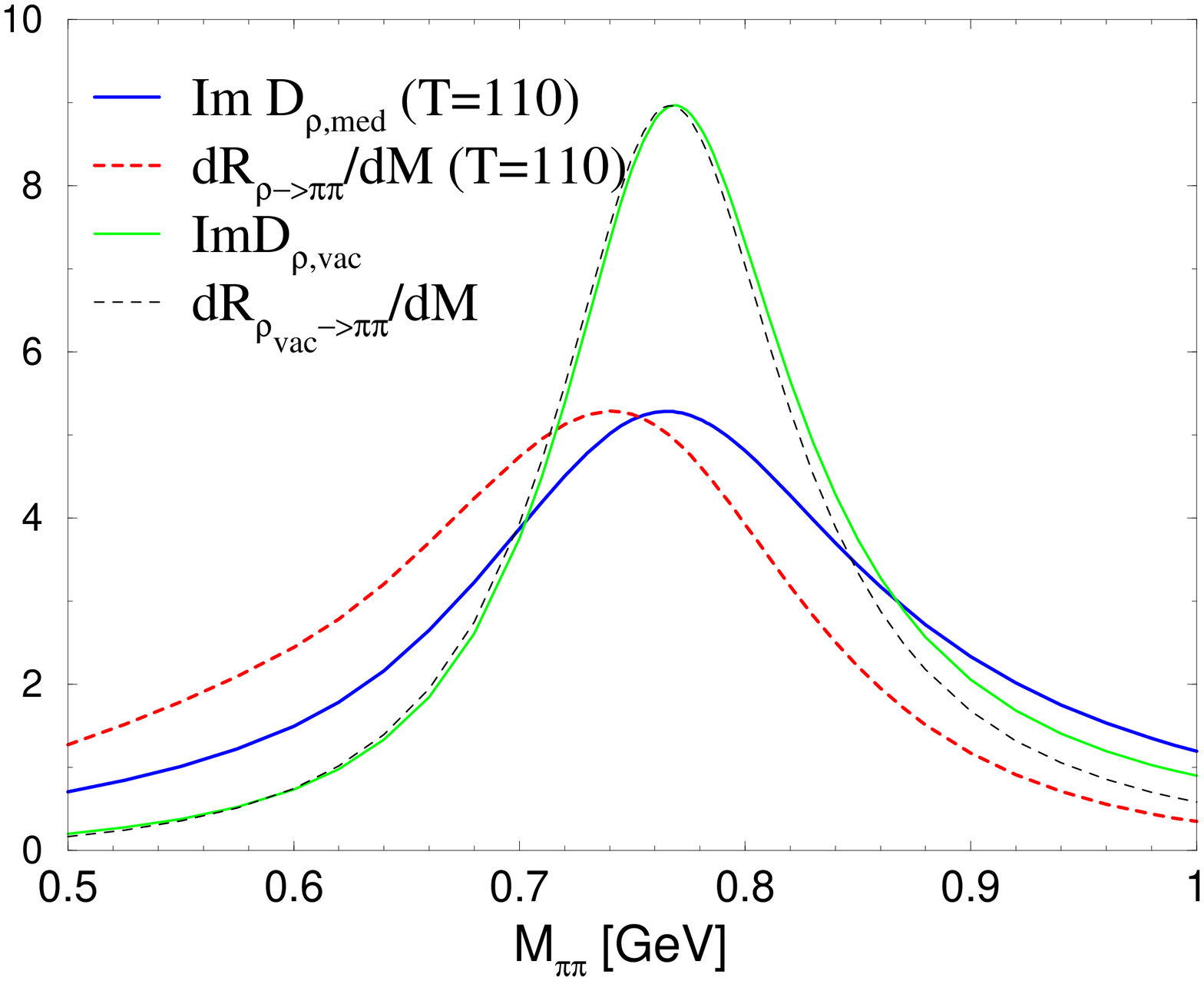,width=7cm,angle=-90}
\ece
\caption{Comparison of $\rho$-meson spectral functions (solid 
lines, in units of GeV$^{-2}$) with invariant-mass distributions for 
$\rho^0\to\pi^+\pi^-$ decay rates (dashed lines, arbitrary units) 
in an environment characteristic for $p$-$p$ (narrow resonance 
curves) and $Au$-$Au$ (broader resonance curves) collisions.
}
\label{fig_rho-stat}
\efig
It is instructive to illustrate the impact of the various 
factors in the  $\pi\pi$ production rate, Eq.(\ref{dRdM}),
relative to the $\rho$ spectral function, for a static environment, 
\ie, at fixed temperature and chemical potentials. To represent the 
$p$-$p$ case, we use the vacuum $\rho$ spectral function and a 
temperature of $T=170$~MeV characteristic for the pertinent hadron 
production systematics. Comparing the $\pi\pi$ invariant-mass 
distribution with the bare spectral function in Fig.~\ref{fig_rho-stat}, 
one observes a slight shift of about -5~MeV, caused by a competition of 
the thermal occupation factor (inducing a downward shift) and the 
various phase space factors (which mostly increase with $M$ favoring an 
upward shift).  For the $Au$-$Au$ case, the appropriate conditions 
correspond to the thermal freezeout stages as extracted, \eg, from blast 
wave fits to hadron $p_t$-spectra. We here adopt the parameters 
determined  by hydrodynamic calculations~\cite{KR02} based on suitably 
constructed chemical-off-equilibrium hadronic trajectories~\cite{Ra02}, 
yielding $T\simeq 110$~MeV, $\mu_\pi=0.5\mu_\rho\simeq 90$~MeV, 
$\mu_N\simeq365$~MeV, etc., translating into a total baryon+antibaryon 
density $\varrho_{B+\bar B}=0.24\varrho_0$. We first note that, under 
these conditions, the $\rho$ spectral function is still appreciably 
broadened, which, after multiplying with phase space and thermal 
factors, results in a $\sim$-25~MeV relative shift of the maximum in 
the $\pi\pi$ invariant-mass distribution, cf.~Fig.~\ref{fig_rho-stat}.     
The rather low temperature also induces a significant shape 
distortion which leads to a more pronounced "apparent" resonance shift, 
in the following sense: when evaluating the shift of the resonance
curves at the full-width half-maximum (FWHM), it amounts to about -45~MeV
on {\em both} sides of the resonance. 

For a more realistic description of the emitted (undistorted) pion
pairs from the late stages  of a heavy-ion collision one should 
allow for a finite emission duration over a profile in 
temperature and density around the expected (average) freezeout 
conditions. A simplified way to do this is using a schematic
thermal fireball model based on cylindrical volume expansion 
with parameters consistent with experiment. The space-time 
integrated emission rate then becomes    
\bea
\frac{dN_{\rho\to\pi\pi}}{dM} &=&
  \frac{3}{\pi^3} \ {\rm Im}\Sigma_{\rho\pi\pi}(M) \ M 
  \int\limits_{t_0}^{\infty} dt \ V_{FB}(t) \ P(t,t_{max}) 
\nonumber\\
& & \times  
\int \frac{q^2 dq}{q_0} \ f^\rho(q_0;T(t),\mu_\rho(t)) \
\nonumber\\
 & &  \times \ {\rm Im} D_\rho (M,q; \mu_B(t),T(t)) \ Acc(M,q) \ ,
\label{dNdM_evo}
\eea
where $V_{FB}(t)$ denotes the time-dependent (isotropic) volume,
$Acc(M,q)$ accounts for experimental acceptance, and the 
function $P(t,t_{max})$ represents the probability distribution  
of $\pi^+\pi^-$ emission around the "average" freezeout. 
For illustration purposes we here assume a Gaussian profile, 
\beq
P(t,t_{max}) = P_0 \  
   \exp\left[(t-t_{max})^2/2\sigma_t^2 \right] \ 
\label{Pem}
\eeq
with $t_{max}$ the time of maximal emission, 
a width characterized by $\sigma_t$ and a (dimensionless)
constant $P_0\le 1$.   
A more detailed treatment with explicit inclusion of pion absorption 
effects will be given elsewhere~\cite{Kolo03}.   
In combination with the increasing fireball volume, $V_{FB}(t)$,
the resulting emission time profile resembles  the example
given in Ref.~\cite{SB03} based on hydrodynamical estimates. 
%\beq
%V_{FC}^{(2)}(t)= 2 \ (z_0+v_z t +\frac{1}{2} a_z t^2) \ \pi \
%(r_0+\frac{1}{2} a_\perp t^2)^2 
%\eeq
The resulting $\pi\pi$ invariant-mass spectra shown in 
Fig.~\ref{fig_rho-evo} are computed with
$\sigma_t=1$~fm/c and $P_0=0.5$ (translating into an effective emission
duration $\Delta t\simeq 1.25$~fm/c), as well as for two average 
freezeout temperatures (with appropriate pion-chemical 
potentials~\cite{Ra02}) characteristic for central $Au$-$Au$ at full 
RHIC energy.  When employing vacuum $\rho$ spectral functions, the peak 
positions are located at about $M_{\pi\pi}\simeq 760$~MeV (\ie, 10~MeV 
below the nominal mass), whereas the in-medium broadening 
effects~\cite{RW99,Ra01} lower the maxima to about 
$M_{\pi\pi}\simeq$~740-745~MeV. Again, the "apparent" mass shift at the
center of the FWHM is larger due to the distorted line 
shape. One can quantify this feature by defining a "centroid" of the mass
distribution as
\beq
\left(m_{\pi\pi}^{cent}\right)^2= \int dM \ M^2 \ \frac{dN}{dM} \quad /  
                  \quad \int dM \ \frac{dN}{dM} \ .
\eeq
For the $\bar T$=110(125)~MeV scenario we find 
$m_{\pi\pi}^{cent}=722(739)$~MeV and 758(769)~MeV with the in-medium 
and free $\rho$ spectral function, respectively.  
Furthermore, note that in both cases, the spectral shapes are relatively 
robust with respect to (w.r.t.) variations in the freezeout temperature, 
whereas the yield is reduced by $\sim$20(15)\% for the free (in-medium) 
spectral function. We will return to the latter issue in 
Sec.~\ref{ssec_ratios}.    
\bfig[htb]
\bce
\epsfig{figure=dNdMppfo.eps,width=8.5cm,angle=0}
\ece
\caption{Time-integrated $\pi^+\pi^-$ emission spectra from thermal 
$\rho^0$-decays in the late stages of central $Au$-$Au$ collisions at 
RHIC based on Eqs.~(\ref{dNdM_evo}) and (\ref{Pem}). The dashed-
dotted (short-dashed) and full (long-dashed) lines are for 
$\bar T=T(t_{max})=110(125)$~MeV using free and in-medium 
$\rho$ spectral functions, respectively.  
}
\label{fig_rho-evo}
\efig

%%%%%%%%%%%%%%%%%%%%%%%%%%%%%%%%%%%%%%%%%%%%%%%%%%%%%%%%%%%%%%%%%%%%%%%
\section{$S$- and $D$-Wave $\pi\pi$ Correlations and Emission} 
\label{sec_SDwave}
%%%%%%%%%%%%%%%%%%%%%%%%%%%%%%%%%%%%%%%%%%%%%%%%%%%%%%%%%%%%%%%%%%%%%%%

%%%%%%%%%%%%%%%%%%%%%%%%%%%%%%%%%%%%%%%%%%%%%%%%%%%%%%%%%%%%%%%%%%%%%%%%
\subsection{Microscopic $\pi\pi$ Interactions and Strength Distributions}
\label{ssec_juel}
%%%%%%%%%%%%%%%%%%%%%%%%%%%%%%%%%%%%%%%%%%%%%%%%%%%%%%%%%%%%%%%%%%%%%%%%

In addition to direct $\rho$-meson decays, a number of other sources of 
(strongly) correlated $\pi^-\pi^+$ emission contribute to their 
invariant-mass distribution. These sources are either due to feeddown 
from resonances (\eg, $a_1\to \rho\pi$) or "direct" $\pi\pi$ 
correlations. The former will be addressed in Sect.~\ref{sec_feed}; 
the latter, which we will focus on now, are determined by the $\pi\pi$ 
interaction strength in a given spin-isospin channel; for an isospin-1 
$P$-wave ($JI$=11) the strength distribution is essentially saturated 
by the $\rho$-meson (\ie, $\rho$ decays), but non-resonant contributions 
are in principle also present. In the $S$-wave, $\pi\pi$ interactions 
are sizable for both isoscalar ($JI$=00) and isotensor ($JI$=02) 
states.  The former represent the well-known "$\sigma$"-channel, with 
ongoing controversy on the nature of the observed resonance structures. 
For the emission strength the relevant quantity is  
the imaginary part of the scattering amplitude, as 
determined via $\pi\pi$ scattering phase shifts~\footnote{This is
strictly correct only as long as no medium effects are included, 
since the latter may be quite sensitive to the underlying
microscopic interactions (cf., \eg, Refs.~\cite{ARCSW95} for a 
detailed analysis in cold nuclear matter).}.   
In the $D$-wave, the only significant feature is the $f_2(1270)$
resonance, with little interaction strength below $M_{\pi\pi}=1$~GeV.  

A microscopic model that accurately reproduces free $\pi\pi$ 
scattering in $S$-, $P$- and $D$-waves up to invariant masses of
about 1.5~GeV, is the J\"ulich $\pi\pi$ interaction~\cite{LDHS90}. 
It is based on an effective Lagrangian generating $s$-, $t$- and 
$u$-channel meson exchanges which are iterated to all orders
by solving an underlying Lippmann-Schwinger equation, 
\beq
{\cal M}^{JI}=V^{JI}+V^{JI} \ G_{\pi\pi} \ {\cal M}^{JI} \ , 
\eeq
for the 
scattering amplitude ${\cal M}^{JI}$; $V^{JI}$ and $G_{\pi\pi}$
are the two-body potentials (Born amplitudes) and 2-pion propagator, 
respectively. In the following we employ a chirally improved 
version~\cite{RDW96} thereof, which has been extended by  
$\pi\pi$ contact interactions
to satisfy constraints from chiral symmetry (these are
necessary for a reliable treatment of in-medium effects).  
In Fig.~\ref{fig_sgrof2}, the solid lines represent the 
imaginary part of the $\pi\pi$ scattering amplitude in the "$\sigma$",
$\rho$ and $f_2$ channels as obtained from the J\"ulich model 
in vacuum, and including medium effects on the two-pion propagator 
within a hot pion gas~\cite{RW93} corresponding to RHIC freezeout 
conditions (direct resonances, such as $\rho\pi\to a_1$, as well as 
baryonic effects are not accounted for in this Section; a more
complete treatment in "$\sigma$" and $f_2$ channels will be reported 
elsewhere).

To obtain the pertinent $\pi^+\pi^-$ emission rates, two further 
steps are necessary. First, recalling Eq.~(\ref{dNdxdq}), the 
combination of spectral function and decay width has to be related 
to $Im {\cal M}$. Using $\Sigma_{\rho\pi\pi}=v_\rho G_{\pi\pi} v_\rho$
and ${\cal M}^{11}=v_\rho D_\rho v_\rho$ ($v_\rho$: $\rho\pi\pi$ 
vertex function), one obtains 
\beq
{\rm Im}D_\rho \ {\rm Im}\Sigma_\rho = {\rm Im}{\cal M}^{11} \  
 {\rm Im}G_{\pi\pi} 
\label{ImM}
\eeq
with ${\rm Im}G_{\pi\pi}=k/(16\pi M)$. 
Second, the isospin basis has to be transformed into particle
basis for the two outgoing pions. For $|II_z\rangle=|10\rangle$ states
(\ie, neutral $\rho$-mesons), the overlap with $\pi^+\pi^-$ pairs 
is 100\%. For $S$- and $D$-waves, however, isospin $I$=0 and
$I$=2 states contribute, 
\beq
{\cal M}^J_{\pi^+\pi^-} = \frac{2}{3} \ {\cal M}^{J,I=0} 
+ \frac{1}{3} {\cal M}^{J,I=2} \ ,  
\eeq 
with only 2/3 of the "$\sigma$"- and $f_2(1270)$-mesons decaying 
into $\pi^+\pi^-$ (the other 1/3 into $\pi^0\pi^0$). 

The 3-momentum integrated emission-strength distributions,
$dR^{JI}/dM$, in the individual spin-isospin channels are shown
by the dashed lines in Fig.~\ref{fig_sgrof2} (in arbitrary 
normalization). Both $\rho$- and
$f_2$-resonance peaks are slightly shifted downward mostly due to 
the thermal occupation factors, between 10-20(30-40)~MeV at 
T=170(110)~MeV. The broad structure in the $\sigma$-channel
is more strongly distorted by the thermal weights, with an 
appreciable shift of strength towards lower masses, especially 
for the smaller temperature.
However, these effects are less pronounced than the ones 
obtained in Refs.~\cite{SB03,KP03}.  This is a consequence of the 
%unrealistic 
schematic $\sigma$ mass distributions   
used in those works which imply an overestimation
of the (free) $\pi\pi$ scattering phase shifts, and thus of
the low invariant-mass strength, in the scalar-isoscalar 
channel.  
\bfig[!t]
\bce
\epsfig{figure=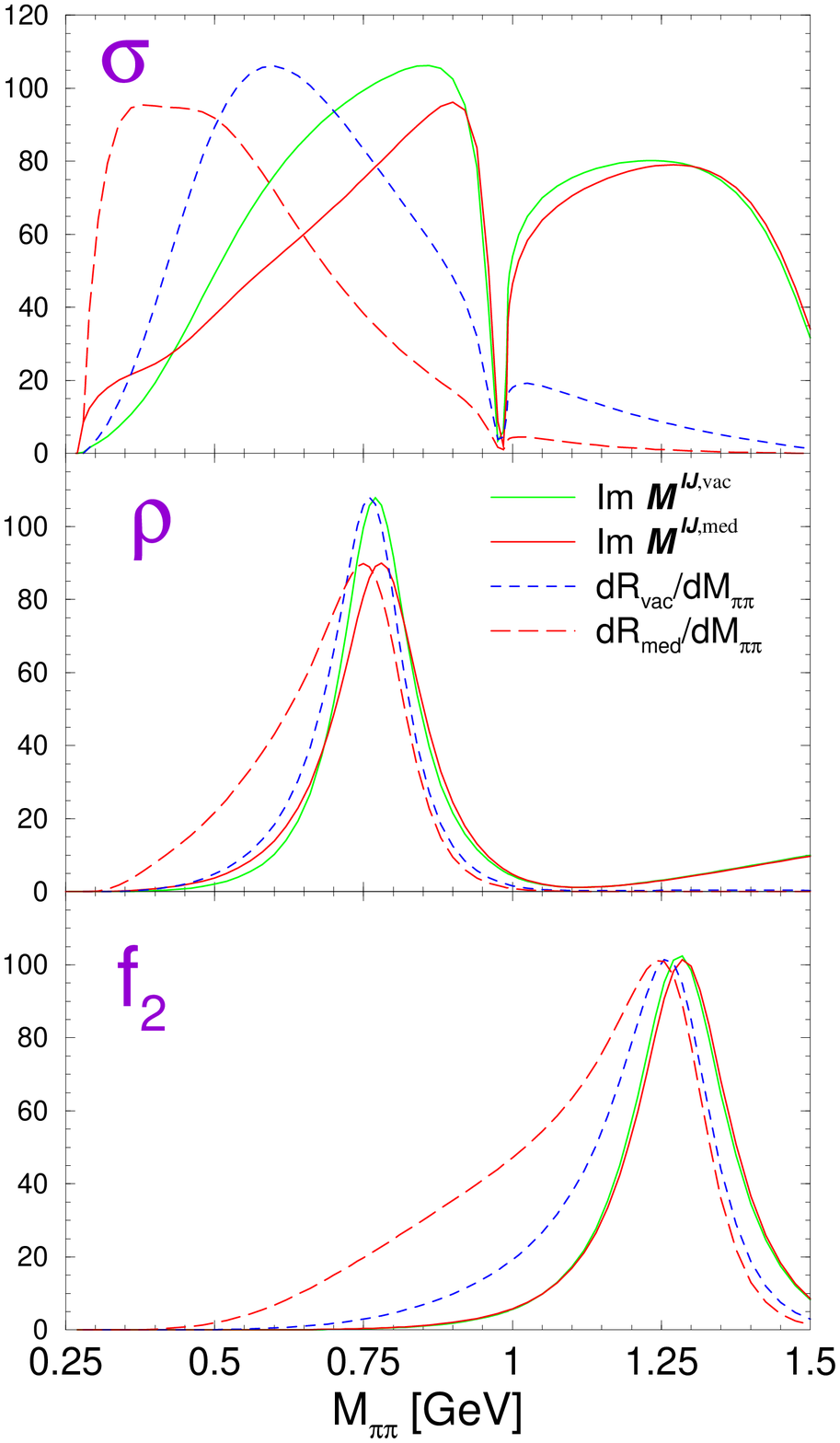,width=8cm,angle=0}
\ece
\caption{Imaginary part of $\pi\pi$ scattering amplitudes within
the J\"ulich model
(solid lines; light: in vacuum, dark: including medium effects at
$T$=110~MeV and $\mu_\pi$=90~MeV) and corresponding $\pi^+\pi^-$
production rates (arbitrarily normalized; short-dashed line:
$T$=170 MeV with vacuum amplitudes, long-dashed line: $T$=110~MeV,
$\mu_\pi$=90~MeV with in-medium amplitudes).  }
\label{fig_sgrof2}
\efig
 
%%%%%%%%%%%%%%%%%%%%%%%%%%%%%%%%%%%%%%%%%%%%%%%%%%%%
\subsection{Thermal $\pi^+\pi^-$ Emission Spectra}
\label{ssec_direct}
%%%%%%%%%%%%%%%%%%%%%%%%%%%%%%%%%%%%%%%%%%%%%%%%%%%%

Implementing the relation (\ref{ImM}) into Eq.~(\ref{dNdxdM}),
the  spin-isospin weighted sum for $\pi^+\pi^-$ thermal emission 
rates can be expressed through the $\pi\pi$ scattering amplitudes 
in $S$-, $P$- and $D$-waves as
\bea
\frac{dR_{\pi^+\pi^-}}{dM}&=& \frac{k}{(2\pi)^4} \int \frac{q^2 dq}{q_0} 
f^B(q_0;2\mu_\pi,T) 
\nonumber\\ 
&& \times 
\sum\limits_J (2J+1) \ {\rm Im}{\cal M}_{\pi^+\pi^-}^J(M;\mu_\pi,T) \ .  
\eea 

\bfig[htb]
\bce
\epsfig{figure=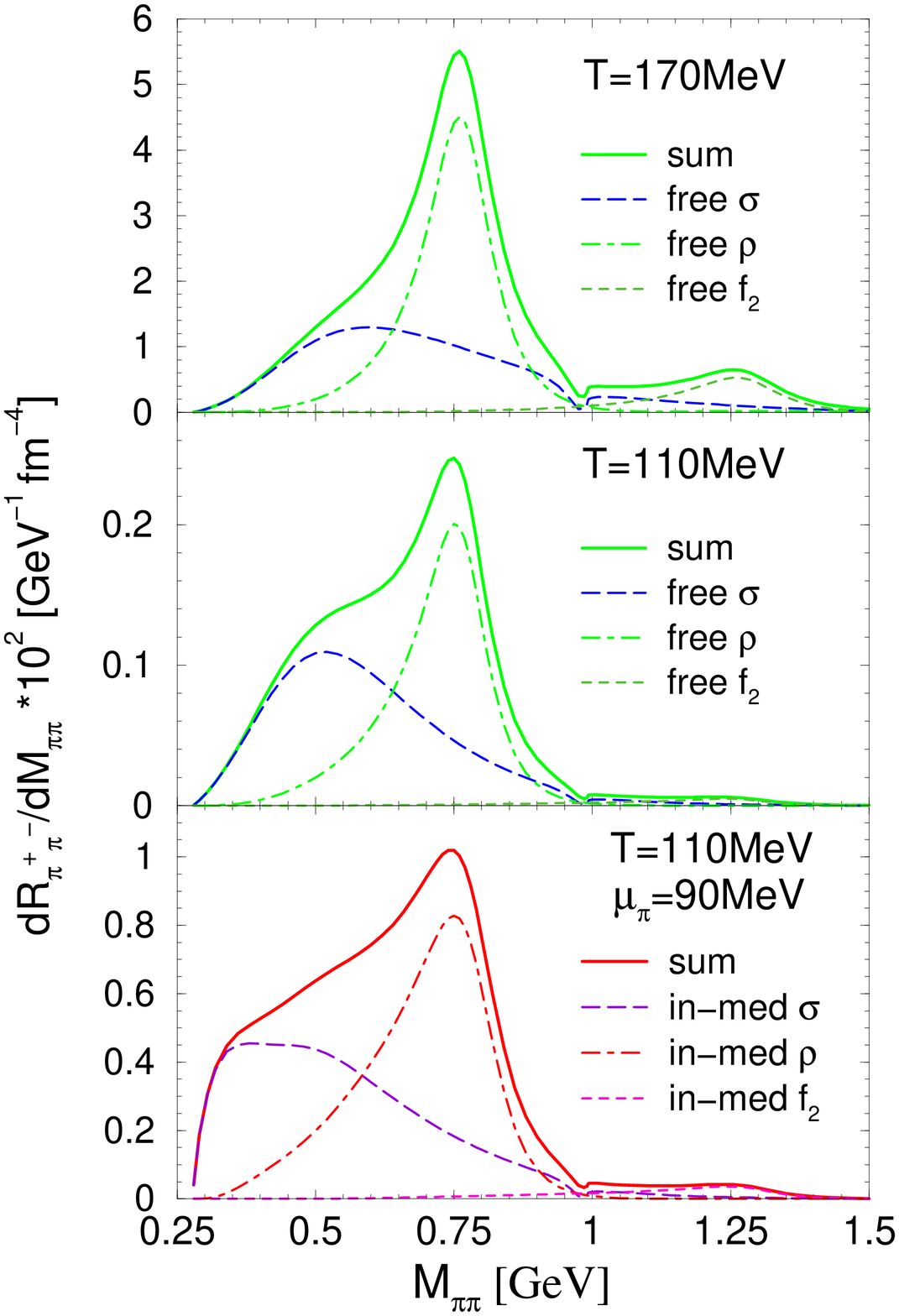,width=8.5cm,angle=0}
\ece
\caption{Thermal $\pi^+\pi^-$ emission rates employing the J\"ulich 
model scattering amplitudes of Fig.~\ref{fig_sgrof2}. Upper and middle 
panel: using free $\pi\pi$ amplitudes at T=170~MeV and T=110~MeV, 
respectively, and $\mu_\pi$=0; lower panel: using in-medium 
amplitudes with finite-$T$ pion modifications~\protect\cite{RW93} 
at T=100~MeV and $\mu_\pi$=90~MeV. 
}
\label{fig_pipi}
\efig
The combined results of the J\"ulich $\pi\pi$ interaction from 
Fig.~\ref{fig_sgrof2} are displayed in Fig.~\ref{fig_pipi} for
temperatures and pion-chemical potentials believed to resemble 
freezeout conditions in $p$-$p$ and central $Au$-$Au$ collisions. 
At T=170~MeV (upper panel), where free $\pi\pi$ amplitudes have been 
employed, the spectrum is dominated by the $\rho$ resonance
with a slight low-mass shoulder from scalar correlations, 
and a clearly discernible $f_2$ resonance. The maximum of the $\rho$
peak is located at $M_{\pi\pi}=760$~MeV. Reducing the emission 
temperature to T=110~MeV, the stronger distortion of the "$\sigma$"  
increases the low-mass shoulder, whereas the relative height of the 
$f_2$ peak is suppressed.  Also note that the $\rho$ 
resonance is situated on the decreasing slope of the "$\sigma$" 
distribution, which entails an additional apparent 5-10~MeV 
downward shift of its position.
The lower panel in Fig.~\ref{fig_pipi} finally incorporates 
finite-$T$ in-medium effects on the pions in the scattering 
amplitudes~\cite{RW93}, as well as finite meson-chemical potentials 
(as inferred from chemistry-conserving thermodynamic 
trajectories~\cite{Ra02}).
One observes a reinforced threshold enhancement, as well
as a net shift of the $\rho$-peak by about -25~MeV,  (recall that 
baryonic medium effects, as well as direct $\rho$-hadron resonances, 
are not included here).  

Another noteworthy feature is the minimum structure in the 
$f_0(980)$ resonance region, which is a direct consequence of the
$180^\circ$-crossing of the scalar-isoscalar $\pi$-$\pi$ phase shifts, 
\ie, the coherence in formation and decay process of the $f_0$ 
(this is different for the "$\sigma$", $\rho$ and $f_2$ resonances,
which induce the usual $90^\circ$-crossing). 
If, on the contrary,  a $\pi\pi$ resonance structure is observed in 
the $f_0(980)$ region (as, \eg, in $pp$ collisions~\cite{star03}), 
it seems to require a different production mechanism, such as 
string breaking. More subtle effects associated with the controversial 
nature of the $f_0(980)$ might also play a role~\cite{SKS03}, 
\eg, formation through the $K\bar K$ channel.     

%%%%%%%%%%%%%%%%%%%%%%%%%%%%%%%%%%%%%%%%%%%%%%%%%%%%
\subsection{Final-State Bose-Einstein Correlations}
\label{ssec_bose} 
%%%%%%%%%%%%%%%%%%%%%%%%%%%%%%%%%%%%%%%%%%%%%%%%%%%%
Apparent mass shifts of $\rho$ resonances detected via the $\pi\pi$
channel are a well-known phenomenon in elementary  $p$-$p$ collisions
and hadronic jets in $e^+e^-$ annihilation at high energy. Among 
the possible explanations~\cite{Laf93} are interferences between direct 
and secondary (through $\pi$-$\pi$ rescattering) $\rho$ production, as 
well as phase space distortions due to Bose-Einstein correlations 
between the decay-pions and the surrounding ones. Whereas the role 
of the former might be reduced in heavy-ion collisions, the opposite 
can be expected for the latter. 
 
%Indeed, even in the elemntary systems tendency towards more
%pronounced effects with higher multiplicity has been observed.  
A naive way of introducing statistical final-state cor\-relations 
(which were not addressed in Refs.~\cite{BBFSB91,SB03,KP03}) into the 
thermal emission rate consists of modifying the free final-state 
$\pi\pi$ decay by an extra Bosefactor, \ie, implementing
a factor $[1+2f^\pi(\omega_k;T)]$ into Im$\Sigma_{\rho\pi\pi}$
in Eq.~(\ref{dNdxdM}). However, this procedure neglects the dependence
on the total three-momentum $q$ of the pion-pair (w.r.t. the thermal 
frame). A better approximation is obtained
by performing an angular average according to~\cite{BBDS92}  
\bea
& & \langle 1+f^\pi(\vec k;T)+f^\pi(-\vec k;T) \rangle =
\nonumber\\
&=& \frac{1}{2} \int\limits_{-1}^{+1} d\cos\theta \ 
[1+f^\pi(\vec k_+';T)+f^\pi(\vec k_-';T)]
\nonumber\\
&=&\frac{T}{\beta \gamma k}  \ln\left\{ \frac
{\sinh\left(\frac{1}{2T} [\gamma (\omega_k+\beta k)-\mu_\pi]\right)}
{\sinh\left(\frac{1}{2T} [\gamma (\omega_k-\beta k)-\mu_\pi]\right)} 
                                 \right\} 
\eea
which depends on the pair 3-momentum and invariant-mass through
the Lorentz factors $\beta$=$q/q_0$ and $\gamma$=$q_0/M_{\pi\pi}$ 
($\vec k_\pm'$ are the pion momenta in the thermal rest frame). 
Consequently, Im$\Sigma_{\rho\pi\pi}$ in Eq.~(\ref{dNdxdM}) has to be 
evaluated inside the momentum integral.  
An upper estimate of the final-state Bose correlations may 
be obtained by inserting the same temperature and chemical potentials
as in the spectral function in Eq.~(\ref{dNdxdM}). This leads to an
appreciable impact on the $\pi^-\pi^-$ emission spectra, 
cf.~Fig.~\ref{fig_bose}: in the $\rho$-resonance region, the rate 
increases by 20~\% ("stimulated" emission), accompanied by a small
downward move (5-10~MeV) of the peak position. Towards the
two-pion threshold, a more pronounced "$\sigma$" shoulder emerges.
The finite-$q$ corrections moderate the Bose enhancement close
to the two-pion threshold, most notably for low temperature and large 
pion-chemical potential.   
\bfig[htb]
\bce
\epsfig{figure=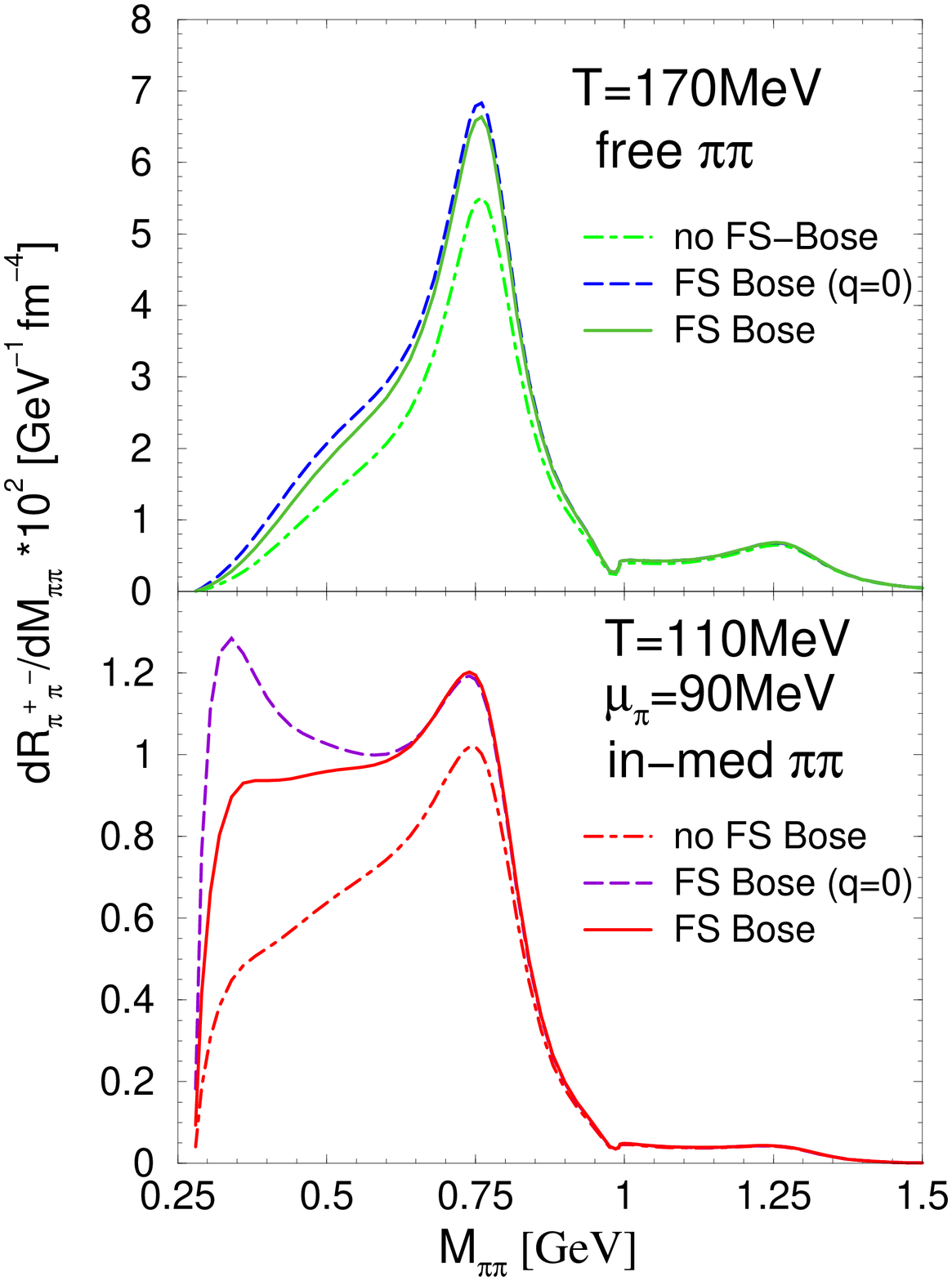,width=8cm,angle=0}
\ece
\caption{Thermal $\pi^+\pi^-$ emission rates employing the J\"ulich
model scattering amplitudes with additional account for final-state 
Bose-Einstein enhancement. Dashed-dotted lines represent the 
results of Sec.~\protect\ref{ssec_direct} (solid lines in 
Fig.~\protect\ref{fig_pipi}), whereas solid and dashed 
lines include the final-state Bosefactors with and without
finite-$q$ corrections, respectively.   
}
\label{fig_bose}
\efig

%%%%%%%%%%%%%%%%%%%%%%%%%%%%%%%%%%%%%%%%%%%%%%%%%%%%%%%%%%%%%%%%%%%%%%%
\section{Resonance Feeddown and Ratios}
\label{sec_feed}
%%%%%%%%%%%%%%%%%%%%%%%%%%%%%%%%%%%%%%%%%%%%%%%%%%%%%%%%%%%%%%%%%%%%%%%
We now return to the question of feeddown contributions to 
$\pi^+\pi^-$ distributions, concerning both their magnitude and 
spectral shape.  
For simplicity, we will focus on the $\rho$ meson, which is 
well-established  in a large number of mesonic and baryonic 
resonance-decay branchings (less so for the scalar "$\sigma$"- and 
tensor $f_2$-mesons). 

%%%%%%%%%%%%%%%%%%%%%%%%%%%%%%%%%%%%%%%%%%%%%%%%%%%%%%%%%%%%%%%%%%%%%%%%
\subsection{Spectral $\pi\pi$ Shape from $a_1$ Decays}
\label{ssec_a1dec}
%%%%%%%%%%%%%%%%%%%%%%%%%%%%%%%%%%%%%%%%%%%%%%%%%%%%%%%%%%%%%%%%%%%%%%%%
As a typical example for resonance feeddown, we here concentrate 
the $a_1(1260) \to \rho\pi$ decay (which gives the largest individual 
contribution). Starting point is the 3-momentum integrated emission 
rate, Eq.~(\ref{dRdM}), which for the present case takes the form  
\bea
\frac{dR_{a_1\to\rho\pi}}{dM_a} &=& \frac{6}{\pi} \
{\rm Im} \Sigma_{a_1\rho\pi}(M_a) \ {\rm Im} D_{a_1}(M_a;\mu_B,T)
\nonumber\\
 & & \times  \left(\frac{M_a T}{2\pi}\right)^{3/2} \
{\rm e}^{-(M_a-\mu_{a_1})/T} \ .
\label{dRdMa}
\eea
The $a_1\rho\pi$ selfenergy is related to the pertinent partial
decay width via 
$-{\rm Im} \Sigma_{a_1\rho\pi}=M_a \Gamma_{a_1\to\rho\pi}$. 
Including the $\rho$-mass distribution through its spectral function
$A_\rho(M)$, one has
\beq
\Gamma_{a_1\to\rho\pi}(M_a) = \frac{1}{12\pi M_a^2} 
\  \int\limits_{2m_\pi}^{M_{max}} \frac{M dM}{\pi} \ 
A_\rho(M) \ q_{cm} \ v_{\rho\pi a_1}^2 \  ,  
\label{gama1}
\eeq
where $q_{cm}$ denotes the three-momentum of the decay products $\rho$ 
(mass $M$) and $\pi$ (mass $m_\pi$) in the $a_1$ rest frame,
$v_{\rho\pi a_1}$ the vertex function, and  $M_{max}=M_a-m_\pi$. 
%For the vertex function $v_{\rho\pi a_1}$,  
%several choices are possible. We here employ the same one as used to 
%compute medium effects in the $\rho$ propagator  
%in Sect.~\ref{ssec_in-med-rho} which reads~\cite{RG99}
%\bea 
%v_{\rho\pi a_1}^2(M,M_a) &=& G_{\rho\pi a_1}^2 \ 
%F_{\rho\pi a_1}(q_{cm})^2 
%\nonumber\\
%& \times & [\frac{1}{2} (M_a^2-M^2-m_\pi^2)^2 +M^2 \omega_\pi(q_{cm})^2 ] 
%  \ . 
%\nonumber\\
%\eea
%From Eq.~(\ref{gama1}) the mass distribution of 
%$\rho$-mesons in $a_1$-decays follows as
%\beq
%\frac{d\Gamma_{a_1\to\rho\pi}}{dM} = \frac{M}{12\pi^2 M_a^2} 
% \ A_\rho(M) \ q_{cm} \ v_{\rho\pi a_1}^2 \  
%\eeq
%which allows to obtain the double-differential $a_1$ decay rate, 
%$dR_{a_1\to\rho\pi}/dM_a dM$. 
Using the differential form of Eq.~(\ref{gama1}), and integrating 
Eq.~(\ref{dRdMa}) over the $a_1$-mass distribution, yields the 
$\rho$-like $\pi^+\pi^-$ production rate from $a_1$ deacys as 
\bea 
\frac{dR_{a_1\to\pi (\pi^+\pi^-)}}{dM} &=& -\frac{6}{\pi}
\int dM_a M_a \int\frac{d^3k}{(2\pi)^3} \ \frac{M_a}{k_0} 
\ \frac{d\Gamma_{a_1\to\rho\pi}}{dM}
\nonumber\\ 
 & \times & f^{a_1}(k_0;\mu_{a_1},T) \ {\rm Im} D_{a_1}(M_a,k;\mu_B,T) 
 \  ,  
\nonumber\\
\label{dRadM}
\eea 
where we reintroduced the explicit integration over the 
$a_1$ Bose function, $f^{a_1}$, with $k_0=(M_a^2+k^2)^{1/2}$. 
An important ingredient in evaluating Eq.~(\ref{dRadM}) is the 
$a_1$ spectral function; we use a microscopic description with a 
$\rho\pi$ selfenergy which gives an accurate description of 
recent axialvector $\tau$-decay data.   

It turns out that the resulting $\pi^+\pi^-$ mass distributions are 
not much affected (downward peak-shifts of 5-10~MeV even
at low temperatures), which is primarily due to the fact
that the {\em free} $\rho$ spectral function enters into the 
expression for the $a_1$-decay width, Eq.~(\ref{gama1}). Broadening 
effects in the underlying $a_1$ spectral function have very little 
impact either.
Final-state Bose enhancement factors, implemented along the lines of 
Sec.~\ref{ssec_bose}, can lead to net peak-shifts of maximally -15~MeV. 
Note that the relative contribution from resonance feeddown is
expected to decrease with decreasing temperatures.

%%%%%%%%%%%%%%%%%%%%%%%%%%%%%%%%%%%%%%%%%%%%%%%%%%%%%%%%%%%%%%%%%%%%%%%%
\subsection{$\rho/\pi$ and $K^*/K$ Ratios}
\label{ssec_ratios}
%%%%%%%%%%%%%%%%%%%%%%%%%%%%%%%%%%%%%%%%%%%%%%%%%%%%%%%%%%%%%%%%%%%%%%%%
For a thermal hadron gas at $T=180$~MeV in full chemical equilibrium, 
an additional $\sim$~60\% of 
the equilibrium $\rho$-number will  emerge from resonance 
feeddown (this excludes subthreshold contributions of the type
$\omega\to\rho\pi$ or $N(1520)\to N\rho$ which do not feed  
into the $\rho$-meson peak); the largest component 
of about 13\% each arises from  
$a_1(1260) \to \rho\pi$ and $a_2(1320) \to \rho\pi$ decays
(at $\mu_B=0$, about 10\% arise from anti-/baryon decays).
The resulting $\rho^0/\pi^-$ ratio ranges between 0.106 and 
0.125 depending on whether weak decays are included in 
the $\pi^-$ yield or not. Such values are in good agreement with 
previous measurements in $e^+e^-$ and $pp$ collisions at 
$\sqrt{s}\simeq 10$-100~GeV. The recent (preliminary) STAR 
measurement for $\sqrt{s}=200$~GeV $pp$ reports 
$\rho^0/\pi^-=0.173\pm 0.026$~\cite{star03}. 
This value, however, does not account for the possibility that
other $\pi^+\pi^-$ sources (most notably "$\sigma$" correlations) may
"contaminate" the $\rho$ resonance region. From the upper panel of 
Fig.~\ref{fig_pipi} one infers that this could lead to a up to 20\%
correction of the $\rho^0$ yield.     

An important question is how the $\rho^0/\pi^-$ ratio develops when 
going to (central) heavy-ion collisions. The $\rho$ meson constitutes 
one of the strongest agents for pion interactions, an thus for 
maintaining thermal equilibrium. This means that 
$\rho \leftrightarrow \pi\pi$ stays in relative chemical equilibrium, 
and therefore $\mu_\rho =2\mu_\pi$. Finite $\mu_\pi$'s build up 
between chemical and thermal freezeout due to effective pion-number 
conservation (and likewise for kaons, etas, and even antibaryons), 
cf.~Ref.~\cite{Ra02} for a hadronic evolution (\ie, thermodynamic 
trajectory) under RHIC conditions.
To obtain the total number of $\rho^0$
and $\pi^-$ in the later stages one again has to include the 
feeddown corrections of all the resonances, but with their respective 
(effective) chemical potentials (\eg, 
$\mu_{a_1}=\mu_\rho+\mu_\pi=3\mu_\pi$, etc.).
Starting the evolution from chemical freezeout at 
$(\mu_\pi,T)=(0,180)$~MeV, the $\rho^0/\pi^-$  ratio is first almost 
stable, being reduced by only 15\% at $(\mu_\pi,T)=(60,140)$~MeV. 
It then starts falling faster being reduced by 30\%
at $(\mu_\pi,T)=(85,120)$~MeV, \ie, $\rho^0/\pi^-=0.075$-0.087, and
to 0.05-0.06 at $(\mu_\pi,T)=(100,100)$~MeV. These  
ranges thus characterize (equilibrium-) conditions close to 
thermal decoupling in central $Au$-$Au$ collisions. The (preliminary) 
STAR measurement in peripheral $Au$-$Au$ lies significantly higher,
at  0.183$\pm$0.027~\cite{star03}. 
Again, $\pi\pi$ $S$-wave contributions in the spectra could lower this 
value.  Also, the actually detected number of $\rho^0\to \pi^+\pi^-$  
decays might be affected by an interplay between (pion-) absorption 
and ($\rho^0$-) regeneration in the vicinity of thermal freezeout. Here, 
transport simulations should be an adequate tool, which find
a $\rho^0/\pi^-$ ratio of around 8\%~\cite{Blei03}.   

Let us briefly comment on a closely related quantity that has recently 
been measured at RHIC energies, namely the $K^{*}(892)/K$ ratio.  For 
200~GeV $p$-$p$ collisions STAR obtained 0.386$\pm$0.029 (preliminary), 
to be compared to a statistical model value of 0.35-0.36 at $T$=180~MeV. 
For 0-10\% (0-14\%) central $Au$-$Au$ at 200~(130)~AGeV, the ratio was
found to decrease to 0.205$\pm$0.033 (0.26$\pm$0.07)~\cite{starK200,starK130}. 
On the one hand, it has been argued that this implies, based on the 
vacuum $K^{*}$ lifetime of about 4~fm/c, an upper limit on the duration 
of the hadronic phase, which follows if one assumes no regeneration. 
If, on the other hand, the reaction $K^*\leftrightarrow K\pi$ is 
assumed to maintain (relative) chemical equilibrium, 
one has $\mu_{K^*}=\mu_\pi+\mu_K$. Along the thermodynamic trajectory 
discussed above one then finds the $K^{*}/K$ ratio to slowly
decrease, passing through values of 0.235-0.295 at temperatures 
$T$=100-120~MeV (and $\mu_{mes}$$>$0). Thus, even complete 
regeneration (which, in the denser stages, could be facilitated 
by in-medium broadening effects) cannot be ruled out by the current 
measurements.   
The same pattern emerges for the $K^{0*}/h^-$ ratio. The preliminary 
STAR value for 0-20\% (0-14\%) central $Au$-$Au$ at 200~(130)~AGeV
is 0.033$\pm$0.004 (0.042$\pm$0.011)~\cite{star03,starK130}. 
For the hadron gas value at chemical freezeout ($T$=180~MeV) we 
find $K^{0*}/h^-$=0.041, in good agreement with the statistical 
model values of Ref.~\cite{BMRS01}.
Around thermal freezeout ($T$=100-120~MeV) the equilibrium ratio has 
decreased to $K^{0*}/h^-$=0.028-0.035.

%%%%%%%%%%%%%%%%%%%%%%%%%%%%%%%%%%%%%%%%%%%%%%%%%%%%%%%%%%%%%%%%%%%%%%%%
\section{Summary and Conclusions}
\label{sec_concl}
%%%%%%%%%%%%%%%%%%%%%%%%%%%%%%%%%%%%%%%%%%%%%%%%%%%%%%%%%%%%%%%%%%%%%%%%
In the present article we have investigated spectral shapes and yields
of $\pi^+\pi^-$ invariant-mass spectra arising from thermal sources 
characteristic for the decoupling stages of high-energy heavy-ion 
collisions. Our analysis is based on realistic $\pi\pi$ interactions 
in $S$-, $P$- and $D$-waves, adjusted to free scattering phase shifts.  

For the $P$-wave, in line with previous studies~\cite{SB03,KP03}, 
thermal phase-space 
factors were found to induce up to 10~MeV downward shifts of the 
$\rho$-peak position when assuming the vacuum resonance profile. 
%However, in-medium modifications of the $\rho$ spectral function due
%to hadronic reinteractions, which have been evaluated earlier in the 
%context of dilepton production, amount to a significant broadening of 
%about 100~MeV even in dilute hadronic matter.  When combined with 
%thermal phase of the $\pi\pi$ decays, a peak-mass reduction of about 
%30~MeV is found. 
However, when using microscopic in-medium $\rho$ spectral functions, 
for which a broadening of about 80-100~MeV in dilute hadronic matter 
has been predicted earlier, the combination with thermal $\pi\pi$ phase 
space leads to a peak-mass reduction of about 30~MeV.  
In fact, low-mass tails of the $\pi\pi$ distribution lower
its centroid by up to another 20~MeV. These findings are more
pronounced than the corresponding ones of Ref.~\cite{KP03}
(based on schematic spectral functions). 
%but do not exhibit  
%peak shifts of up to -70~MeV as extracted from recent STAR 
%data~\cite{star03} for (peripheral) $Au$-$Au$ collisions.
Although the effects of in-medium modifications qualitatively reflect 
the relative changes between $p$-$p$ and $Au$-$Au$ collisions as 
extracted from preliminary STAR data~\cite{star03}, the experimentally 
observed absolute peak shifts of about -40~MeV and -70~MeV, 
respectively, are not reproduced  by our results.   
%but do not exhibit  
%peak shifts of up to -70~MeV as extracted from recent STAR 
%data~\cite{star03} for (peripheral) $Au$-$Au$ collisions. 

For $S$-wave emission, especially in the scalar-isoscalar 
("$\sigma$") channel, we emphasized the importance of employing 
realistic $\pi\pi$ scattering amplitudes (strength distributions), 
which sensitively respond to distortions by thermal occupation 
factors close to freezeout. This is mandatory for a reliable 
assessment of in-medium effects on the "$\sigma$", which, as for 
the $\rho$, can be related to partial chiral symmetry restoration. 
We found that thermal medium modifications and pion Bose-enhancement 
lead to an extended threshold maximum ($M_{\pi\pi}$=300-500~MeV) in 
the $S$-wave $\pi^+\pi^-$ emission rates. However, its magnitude 
relative to the $P$-wave contribution is moderate, 
translating into a low-mass shoulder of the $\rho$ resonance.  
In addition, the decreasing "$\sigma$" slope in the $\rho$-resonance 
region induces another -10~MeV apparent shift of the observed $\rho$ 
peak. 
The general trend in the preliminary STAR data which indicates a 
decrease in the height of the $\rho$-peak relative to the 
low-mass continuum when going from $p$-$p$ to $Au$-$Au$ collisions, 
is qualitatively reproduced by our results. To what extent this 
reflects a $\rho$ broadening, and what {\em additional} mass 
shift is required, should be evaluated by a more quantitative 
comparison to experimental spectra.  

Finally, we have assessed total $\rho^0$ yields, including feeddown 
corrections. We found that the hadro-chemical freezeout value for 
the $\rho^0/\pi^-$ ratio ($\sim$11\% at $T$=180~MeV) decreases 
by only about 30-40\% towards thermal freezeout ($T$=110-120~MeV),
due to large pion-chemical potentials.   

For future investigations, a more complete treatment of in-medium
effects on the "$\sigma$", combining finite temperatures and 
densities, is desirable. The $f_0(980)\to \pi^+\pi^-$ contribution 
clearly deserves further study. For a realistic comparison with 
experimental data, Dalitz decays (\eg, $\omega \to (\pi^+\pi^-)\pi^0$)
have to be included. Also, pion absorption effects need to be 
addressed explicitly. In addition, the relation to resonance 
modifications in high-energy $pp$ and $e^+e^-$ collisions ought  
to be understood better~\cite{Laf93}. 
Only an overall consistent treatment will eventually allow
for a reliable extraction of in-medium effects from short-lived 
resonance spectroscopy in the heavy-ion environment.

\vskip1cm
 
\centerline {\bf ACKNOWLEDGMENTS}
I thank G.E.~Brown, P.~Fachini and Z.~Xu for interesting discussions, 
and P. Fachini for bringing Ref.~\cite{Laf93} to my attention.

\end{document}